\documentclass{elsart}
\usepackage{bm}

\def\max{{\rm max}}
\def\min{{\rm min}}

\begin{document}

\begin{frontmatter}

\title{On the Dual Cascade in Two-Dimensional Turbulence}

\author{Chuong V. Tran and John C. Bowman}

\address{Department of Mathematical and Statistical Sciences, University 
of Alberta, Edmonton, Alberta, Canada, T6G 2G1.}

\ead{chuong@math.ualberta.ca, bowman@math.ualberta.ca}

\begin{abstract}
We study the dual cascade scenario for two-dimensional turbulence
driven by a spectrally localized forcing applied over a finite
wavenumber range $[k_\min,k_\max]$ (with $k_\min > 0$) such that the
respective energy and enstrophy injection rates $\epsilon$ and $\eta$  
satisfy $k_\min^2\epsilon\le\eta\le k_\max^2\epsilon$.  
The classical Kraichnan--Leith--Batchelor para\-digm, based on
the simultaneous conservation of energy and enstrophy and
the scale-selectivity of the molecular viscosity, requires that
the domain be unbounded in both directions. For two-dimensional
turbulence either in a doubly periodic domain or in an unbounded
channel with a periodic boundary condition in the across-channel
direction, a direct enstrophy cascade is not possible. In the usual
case where the forcing wavenumber is no greater than the geometric
mean of the integral and dissipation wavenumbers, constant spectral
slopes must satisfy $\beta>5$ and $\alpha+\beta\ge8$, where $-\alpha$
($-\beta$) is the asymptotic slope of the range of wavenumbers lower
(higher) than the forcing wavenumber. The influence of a large-scale
dissipation on the realizability of a dual cascade is analyzed. We
discuss the consequences for numerical simulations attempting to mimic
the classical unbounded picture in a bounded domain.

\end{abstract}


\end{frontmatter}


\section{Introduction}

Since Kraichnan \cite{Kraichnan67,Kraichnan71}, Batchelor 
\cite{Batchelor69}, and Leith \cite{Leith68} (referred to as KLB) 
adapted Kolmogorov's theory of self-similarity in three-dimensional
turbulence to two-dimensional (2D) fluids, the conventional
wisdom for decades has been that 2D turbulence simultaneously exhibits a
direct cascade of enstrophy to large wavenumbers, up to a dissipation
wavenumber $k_\nu$, and an inverse cascade of energy to small
wavenumbers, down to wavenumber $k=0$. In the limit of small
viscosity, the inverse cascade is thought to proceed indefinitely in
time to ever-larger scales, transferring virtually all of the energy  
input to wavenumber zero. The direct cascade is thought to come 
into balance with viscosity, transferring virtually all of the enstrophy
input to $k_\nu$, where it will then be dissipated. In the long-time
limit, a quasi-steady state is reached, in which two inertial ranges are
established. According to the KLB theory, the energy range, which is only
quasi-steady, scales as $k^{-5/3}$. The enstrophy range,
which is in absolute equilibrium, should scale as $k^{-3}$. 

The idea of a dual cascade was first suggested by Fj{\o}rtoft
\cite{Fjortoft53}, who examined nonlinear transfer by individual
interacting wavenumber triads. A later study by Merilees and Warn
\cite{Merilees75} provides more quantitative detail. They showed
that roughly 70\% (60\%) of triads containing a given intermediate
wavenumber $k$ predominantly exchange energy (enstrophy) with lower
(higher) wavenumbers. These analyses of nonlinear transfer, although
carried out at the triad level, provide only a necessary basis for the
KLB theory. The theory requires all triads to work collectively such
that virtually all of the energy (enstrophy) input gets
transferred to the large (small) scales; this is neither suggested nor
implied by \cite{Fjortoft53,Merilees75}. Nevertheless, a dual cascade
from a spectrally localized initial spectrum is consistent with
\cite{Fjortoft53,Merilees75} and has been well confirmed by numerical
simulations with various resolutions (e.g. Borue \cite{Borue93,Borue94};
Frisch and Sulem \cite{Frisch84}; Lilly \cite{Lilly71}; Smith and
Yakhot \cite{Smith93}). In particular, the inverse energy cascade is
observed in the laboratory experiments of Dubos {\it et al.\/} \cite{dubos01}
and Paret and Tabeling \cite{Paret97,Paret98}.\footnote{The inverse energy
cascade and the $k^{-5/3}$ range are also seen to be robust
in many numerical simulations, at least until the energy
reaches the lower spectral boundary; however, the evidence for a direct
enstrophy cascade is inconclusive (see the discussion in Paret and Tabeling
\cite{Paret98} and references therein).} What has not been established
beyond doubt is the realization of the inertial spectral scaling
$k^{-3}$. In fact, there exist other theories that propose very
different spectral slopes (Moffatt \cite{Moffatt86}; Saffman
\cite{Saffman71}; Sulem and Frisch \cite{Sulem75}).   

Numerical simulations, aiming to verify the  KLB picture, face a
number of formidable tasks. First, the simulations are performed for
fluids in a rectangular box instead of an infinite domain. This turns out
to be a serious shortcoming, as the equilibrium dynamics of a fluid in a
doubly periodic domain differs considerably from that of an unbounded fluid
in the KLB picture (see Constantin, Foias, and Manley
\cite{Constantin94}; Tran and Shepherd \cite{Tran02a}). In particular,
the~$k^{-3}$ range and the direct enstrophy cascade have been shown to
be unrealizable in a bounded domain. The main reason seems to be that
in the bounded case, where an absolute equilibrium will be reached,
there is no analogue of a persistent upscale flow of energy that
eventually evades viscous dissipation altogether. To mimic the inverse
cascade, one needs to introduce a dissipation that removes energy at
the large scales (cf. \cite{Sukoriansky99}). Linear Ekman drag
(proportional to $k^0$ and restricted to a few low wavenumbers) and
inverse viscosity ($\nu_\mu k^{2\mu}$, with $\mu<0$) have often been
used for that purpose. Unfortunately, the inverse energy cascade,
subject to this large-scale dissipation, carries with it a significant
fraction of the enstrophy (in contrast to the asymptotic KLB inverse
cascade, which carries no enstrophy). Moreover, there is no guarantee
that the large-scale dissipation will absorb virtually all of the
energy input. This is crucial in the KLB picture, as any fraction of
the energy input that gets reflected would ultimately be trapped in
the inertial ranges. The trapped energy, being in a virtually
inviscid region, would then considerably change the dynamics and the
slopes of the inertial ranges. Second, testing the theory requires
the achievement of high Reynolds numbers, but current computers are
only able to resolve relatively low Reynolds numbers. To overcome
these resolution limitations, researchers often resort to introducing
a hyperviscosity $\nu_\mu k^{2\mu}$, where the degree of viscosity
$\mu$ is greater than one. This numerical device helps to compress the
dissipation range, allowing simulations to be performed at relatively
low resolutions. It is hoped that the effect of this modification on
the inertial-range dynamics is negligible. However, a significant
dissipation of energy at the small scales is inevitable in all
numerical schemes. 

In this study, we establish a theoretical basis highlighting the
dynamical differences between bounded and unbounded 2D turbulence.
The consequences for numerical simulations that aim to verify the KLB
theory will be addressed. In particular, we argue that instead of the
familiar inertial-range spectral scalings~$k^{-5/3}$ and~$k^{-3}$
conjectured by KLB, one would expect steeper
scalings, such as~$k^{-3}$ and~$k^{-5}$, respectively, in these
ranges for bounded systems in equilibrium. In fact, it is shown in
\cite{Tran02a} that the enstrophy range must be steeper than $k^{-5}$;
the physical implication is that no direct
enstrophy cascade is possible. This result is easily generalized,
using the Poincar\'e inequality, to turbulence in semi-unbounded 2D
fluids, i.e. fluids confined to unbounded channels with a periodic boundary
condition in the across-channel direction. Moreover, we investigate
the effectiveness of using a large-scale dissipation to obtain a
dual cascade in a bounded fluid. This information should be particularly
useful for numerical simulations.

In Section~2, we summarize the KLB theory and contrast it to the dynamics
of a bounded fluid. We also show that the theory does not apply to a
fluid in an infinite channel with a periodic boundary condition in the
across-channel direction. In Section 3, we review a result in 
\cite{Tran02a} related to the unrealizability of a direct enstrophy
cascade and discuss the fundamental differences between the dynamics
of bounded fluids in equilibrium and that of unbounded fluids (or
bounded fluids in transient phase) in the KLB picture. In Section 4,
we derive a new constraint on the spectral slope of the energy range
for bounded fluids in equilibrium and a condition for a persistent
inverse energy cascade for unbounded fluids. In Section 5, we point
out some effects of including a large-scale dissipation and the
implications for numerical simulations attempting to verify the KLB
picture. We conclude with some remarks in the final section. Further
estimates on spectral slopes are given in the appendices.

\section{KLB picture for unbounded 2D fluids}

The evolution of the ensemble-averaged energy spectrum $E(k)$, which
represents the energy density associated with the wavenumber $k$, is
governed by (see Frisch \cite{Frisch95} and Kraichnan \cite{Kraichnan67}) 
\begin{eqnarray}
\label{E(k)evolution}
\frac{d}{dt}E(k) &=& T(k)-2\nu k^2E(k)+F(k).
\end{eqnarray}
Here $T(k)$ and $F(k)$ are, respectively, the 
ensemble-averaged energy transfer and energy input rate and $\nu$ is 
the kinematic viscosity coefficient. Since waves of the same
scale do not nonlinearly interact,
$T(k)$ is linear in the modal component corresponding to wavenumber
$k$. Moreover, $T(k)$ satisfies, by virtue of energy and
enstrophy conservation,
\begin{eqnarray}
\label{conservation}
\int_0^{\infty}T(k)\,dk &=& \int_0^{\infty}k^2T(k)\,dk = 0.
\end{eqnarray}
For a fluid in a doubly periodic domain, which for convenience we call
a bounded fluid, the integral in (\ref{conservation}) and
elsewhere in this section should be replaced by a discrete sum over
all wavenumbers. This constraint on $T(k)$ imposes certain
restrictions on its distribution and is thought to give rise to the
dual cascade, believed to be a distinct feature of 2D turbulence.  

We multiply (\ref{E(k)evolution}) by $k^2$ and integrate both the
original and resulting equations over all wavenumbers, noting from
(\ref{conservation}) that the nonlinear terms drop out, to obtain
evolution equations for the total energy density $E=\int_0^\infty
E(k)\,dk$ and enstrophy density $Z=\int_0^\infty k^2E(k)\,dk$,
\begin{eqnarray}
\label{Eevolution}
\frac{d}{dt}E &=& -2\nu Z + \epsilon,\\
\label{Zevolution}
\frac{d}{dt}Z &=& -2\nu P + \eta,
\end{eqnarray}
in terms of the energy and enstrophy injection rates 
$\epsilon=\int_0^\infty F(k)\,dk$ and $\eta=\int_0^\infty k^2F(k)\,dk$, 
respectively, and the palinstrophy density $P=\int_0^\infty k^4E(k)\,dk$.
We assume that the forcing is spectrally localized to a wavenumber interval
$[k_\min,k_\max]$ in the sense that
\begin{eqnarray}
\label{localized}
0\le k_\min^2\epsilon\le\eta\le k_\max^2\epsilon.
\end{eqnarray}
This hypothesis is employed in \cite{Tran02a}.
This is a classical (although not exclusive)
scenario for the KLB theory (cf. Kraichnan \cite{Kraichnan67},
p.~1421b; Pouquet et al. \cite{Pouquet75}, p.~314; and Lesieur
\cite{Lesieur97}, p.~291), and is furthermore a common setup in
numerical simulations of forced 2D turbulence (cf. Lilly
\cite{Lilly72}; Basdevant et al. \cite{Basdevant81}; Shepherd
\cite{Shepherd87}). This assumption seems plausible for
time-independent or white-noise forcing over $[k_\min,k_\max]$, given 
the assemble-averaged nature of~$\epsilon$ and~$\eta$. In particular, a 
monoscale forcing at a wavenumber $s$ satisfies~(\ref{localized})
with $k_\min=k_\max=s$ for each individual realization.
Another example of such a forcing over $[k_\min,k_\max]$ is described in
\cite{Tran02a}; in each realization it yields time-independent energy and
enstrophy injection rates $\epsilon$ and $\eta$ such that
$\eta=s^2\epsilon$, where $s^2$ is the mean of $k^2$ over
$[k_\min,k_\max]$. A similar forcing was used by Shepherd
\cite{Shepherd87} in a study of 2D turbulence in a large-scale zonal
jet on the so-called beta-plane. In these examples, the characteristic
forcing wavenumber $s$ is constant in time.

The dual cascade scenario can be best appreciated if one examines the
evolution equation
\begin{eqnarray}
\label{s2E-Z}
s^2\frac{d}{dt}E-\frac{d}{dt}Z &=& 2\nu(P-s^2Z),
\end{eqnarray}
obtained from (\ref{Eevolution}) and (\ref{Zevolution}).
Here $s$ is the forcing wavenumber $\sqrt{\eta/\epsilon}$
(for $\epsilon>0$), which according to hypothesis (\ref{localized}),
must lie in the interval $[k_\min,k_\max]$. If $\epsilon=0$, then
$\eta=0$ and we take $s$ to be any wavenumber in $[k_\min,k_\max]$. A
direct enstrophy cascade requires that the characteristic enstrophy
dissipation wavenumber~$\sqrt{P/Z}$ be much larger than the forcing
wavenumber $s$; hence, the right-hand side of (\ref{s2E-Z}) must be
positive. The positiveness of $P-s^2Z$ implies the positiveness of the
left-hand side of~(\ref{s2E-Z}) as well. If, in accord with the
quasi-steady KLB theory, the total enstrophy reaches a steady state,
we deduce that $dE/dt$ must be positive. Equation~(\ref{s2E-Z})
reflects the fact that the total energy must increase without limit,
due to the inverse energy cascade toward wavenumber zero. (In a
bounded domain, Tran and Shepherd~\cite{Tran02a} showed that $P=s^2Z$
and concluded from this that no direct cascade is possible.) 

Equation (\ref{s2E-Z}) satisfies $P/Z\gg s^2$ not only for the KLB
scaling~$k^{-5/3}$ and~$k^{-3}$ but also for a rich variety of
spectra. This condition only requires an energy spectrum shallower
than $k^{-5}$ for a sufficiently wide range of wavenumbers $k>s$,
provided that the energy spectrum for $k\ll s$ is shallower
than~$k^{-3}$. (In Section~\ref{constraints}, it is shown that the
quantity $P-s^2Z$ can be positive even if the energy spectrum for
$k>s$ is steeper than $k^{-5}$. This allows for the possibility of an
inverse energy cascade in the absence of a direct enstrophy cascade.)
However, the KLB theory insists on the specific scalings $k^{-5/3}$
and $k^{-3}$ (with a logarithmic correction proposed by Kraichnan
\cite{Kraichnan67,Kraichnan71} and further investigated by Bowman
\cite{Bowman96}), respectively, for the energy and enstrophy inertial
ranges. The $k^{-5/3}$ scaling is analogous to 
the Kolmogorov spectrum for 3D turbulence; the $k^{-3}$ scaling implies
that successive octaves in the enstrophy range contain equal amounts
of enstrophy, so that the enstrophy grows logarithmically with
dissipation wavenumber $k_\nu$. There exist a
number of predictions for the numerical value of the enstrophy
inertial range slope $-\beta$ in the literature. Saffman \cite{Saffman71}
proposes $\beta=4$, while Moffat \cite{Moffatt86} favours a slightly
smaller value: $\beta=11/3$. Sulem and Frisch \cite{Sulem75} instead
propose the upper bound $\beta\le11/3$.

In the long-time limit, the KLB inverse energy cascade 
(or any inverse energy cascade with a spectral scaling shallower than
$k^{-3}$ near $k=0$) carries no enstrophy with it. Therefore, the
enstrophy necessarily approaches an absolute equilibrium in a
quasi-steady state. As a consequence, (\ref{s2E-Z}) reduces to   
\begin{eqnarray}
s^2\frac{d}{dt}E &=& 2\nu(P-s^2Z).
\end{eqnarray} 
This equation indicates a simple and interesting fact about the KLB theory
for unbounded 2D turbulence: in a quasi-steady state (for which
$dZ/dt=0$), the strength of an inverse energy cascade (the rate of the
energy growth), if realizable, is primarily determined by the rate $2\nu P$
of enstrophy dissipation.

Besides the simultaneous conservation of energy and enstrophy, other 
essential features of 2D turbulence that underly the KLB theory are
the scale-selectivity of the molecular viscosity and the unboundedness
of the domain (in both directions). Together, they give rise to an
infinite reservoir of energy in the vicinity of $k=0$ that allows for
the possibility of the KLB inverse energy cascade (which contains
no enstrophy if the spectrum near $k=0$ is shallower than
$k^{-3}$). The theory breaks down when either the scale-selectivity of 
the dissipation or the unboundedness of the domain is absent. We
demonstrate the former case and the semi-bounded case below.
The case of a fluid in a doubly periodic domain is studied in
\cite{Tran02a} and will be reviewed in the next section.  

Consider (\ref{E(k)evolution}) with the viscous dissipation term
$\nu k^2$ replaced by a constant $\sigma>0$; this scale-neutral 
frictional dissipation is often called Ekman drag in
geophysical contexts. Equations (\ref{Eevolution}) and
(\ref{Zevolution}) become 
\begin{eqnarray}
\frac{d}{dt}E &=& -2\sigma E + \epsilon,\\
\frac{d}{dt}Z &=& -2\sigma Z + \eta,
\end{eqnarray}
which, for bounded injection rates $\epsilon$ and $\eta$,
implies that both the energy and enstrophy are bounded.
This simple
fact precludes the KLB type of inverse energy cascade. (As argued
below, an inverse energy cascade dissipated by the friction at
the large scales is not possible either.) Moreover, upon
applying (\ref{localized}), we find
\begin{eqnarray}
\frac{d}{dt}(k_\min^2E-Z) &\le& -2\sigma(k_\min^2E-Z),\\
\frac{d}{dt}(k_\max^2E-Z) &\ge& -2\sigma(k_\max^2E-Z).
\end{eqnarray}
Hence, in the limit $t\rightarrow\infty$ the following holds
\begin{eqnarray}
k_\min^2E-Z \le 0 \le k_\max^2E-Z,
\end{eqnarray}
or equivalently,
\begin{eqnarray}
\label{sigma}
k_\min^2 \le \frac{Z}{E} \le k_\max^2.
\end{eqnarray}
Equation (\ref{sigma}) implies that the redistribution of energy and
enstrophy obeys exactly the same constraint as that imposed on the
energy and enstrophy injection rates. Now, the boundedness of energy
(enstrophy) prohibits an infinitely wide range of wavenumbers $k<k_\min$ 
($k>k_\max$) in which the energy spectrum can scale as $k^{-1}$
($k^{-3}$), as this would imply a logarithmic divergence of the energy
(enstrophy) as $k\rightarrow 0$ ($k\rightarrow\infty$). In fact, an
energy spectrum~$k^{-1}$ ($k^{-3}$) or steeper (shallower) on the
large (small) scales is inconsistent with~(\ref{sigma}). Spectra
consistent with (\ref{sigma}) require that the energy and enstrophy be
primarily dissipated near the region of forcing; hence no inverse
(direct) energy (enstrophy) cascade is possible.

Besides the boundedness of the energy density in the present case, the
spectral distribution of energy and enstrophy obeying (\ref{sigma}) is
profoundly different from that of the classical picture. The energy
range is seen to be much shallower than $k^{-5/3}$ for a
friction that acts uniformly on all scales. What seems curious
is that the slopes of the enstrophy range in both cases do not differ
by much. In fact, one could argue that a $k^{-3}$ enstrophy-range spectrum 
with a logarithmic correction is consistent with the constraint
(\ref{sigma}), provided that the spectrum in the energy range scales
as $k^{-1}$ (with a similar correction). Nevertheless, we
see that the scale-selectivity of the 
molecular viscosity plays an important role in the dual cascade
picture. Perhaps, instead of studying the 2D Navier--Stokes dual cascade,
one could replace the molecular dissipation
$\nu k^2$ in (\ref{E(k)evolution}) by another scale-selective
dissipation $\nu_\mu k^{2\mu}$ (with $\mu>0$) and examine
the realizability of a dual cascade in this hypothetical unbounded system.  

Finally, it is interesting to note that the KLB picture is not
realizable for a 2D Navier--Stokes fluid confined to an unbounded channel
with a periodic boundary condition in the across-channel
direction ($0\le y\le L$) and vanishing velocity (and derivatives thereof)
at $x=\pm\infty$. This system is also furnished with the zero-mean flow
condition
\begin{eqnarray}
\int_0^L \bm u(x,y,t)\,dy &=& \bm 0,
\end{eqnarray} 
where $\bm u(x,y,t)$ is the fluid velocity.
To see why the dynamics of this system are incompatible
with the KLB picture, consider the Poincar\'e inequality for this domain:
there exists a constant $\lambda>0$ such that 
\begin{eqnarray}
\label{poincare}
\lambda\int_0^{\infty}k^{\mu}E(k)\,dk &\le& 
\int_0^{\infty}k^{2+\mu}E(k)\,dk,
\end{eqnarray}   
where $\mu\ge0$. The optimal value of the constant $\lambda$ depends
on $L$. Now, the energy equation (\ref{Eevolution}) and
enstrophy equation (\ref{Zevolution}) can be bounded {\it via\/}
(\ref{poincare}) as follows  
\begin{eqnarray}
\frac{d}{dt}E &\le& -2\nu\lambda E + \epsilon,\\
\frac{d}{dt}Z &\le& -2\nu\lambda Z + \eta.
\end{eqnarray}
These inequalities render the boundedness of both energy and
enstrophy. The boundedness of the energy rules out
the persistent inverse cascade needed for the positiveness of 
$P-s^2Z$; rather, an absolute equilibrium is more plausible for
this system. This result is quite physically reasonable:
if one visualizes the inverse cascade as a result of the
coalescence of same-sign vortices to form ever-larger ones,
the system will tend toward equilibrium as the radii of the vortices
approach the channel width $L$. This dynamical behaviour is observed
by Rutgers \cite{Rutgers98} in an experiment of turbulence in a long
channel, where, as equilibrium is approached, the vortices grow until the
channel width is reached.

Now, if an equilibrium is achieved, (\ref{s2E-Z}) necessarily
reduces to   
\begin{eqnarray}
\label{equilibrium}
P-s^2Z &=& 0.
\end{eqnarray}
This equation, which also applies to the case of bounded fluids in
equilibrium considered in the next section, implies that the
dissipation of enstrophy mainly occurs in the forcing region and
admits much steeper spectra than the KLB spectrum.      

{\bf Remark 1.} Some theoretical studies
of the 2D Navier--Stokes equations in unbounded domains make an 
{\it a priori\/} assumption that the Poincar\'e inequality holds (e.g. Rosa
\cite{Rosa98}; see also Temam \cite{Temam97}, p. 307). Such studies
automatically exclude the possibility of the KLB dynamics. 

\section{Dynamics of bounded fluids in equilibrium}

In bounded systems, the energy is also in absolute equilibrium
for arbitrary (but positive) viscosity coefficient $\nu$, so that
(\ref{equilibrium}) holds in equilibrium. This implies that $P/Z=s^2$,
ruling out the existence of an enstrophy inertial range, as argued
in \cite{Tran02a}. Moreover, the energy spectrum for $k>k_\max$ is
steeper than~$k^{-5}$. This constraint clearly indicates a dramatic
departure from the KLB theory. It is consistent with numerous
numerical results, in which large-scale vortices, known as coherent
structures, are observed (see for example Borue \cite{Borue94};
McWilliams \cite{McWilliams84,McWilliams90}; Santangelo, Benzi, and
Legras \cite{Santangelo89}; Smith and Yakhot \cite{Smith93,Smith94});
these are often blamed for causing spectra steeper than those
predicted by KLB. Although the mechanism behind these structures is
not fully understood, we argue that it is the steepness of the
spectrum (steeper than~$k^{-5}$) that allows coherent structures to
form, rather than the other way around. There is no need to invoke
coherent structures to explain steep spectra: the steepness arises merely as a
consequence of global conservation laws, molecular viscosity, and a
spectrally localized forcing. As a matter of fact, a small-scale spectrum
steeper than~$k^{-5}$ implies that the large scales carry
virtually all of the system's enstrophy. Hence, dynamics exhibiting strong
large-scale structures on a much weaker turbulent background of noise
are consistent with \cite{Tran02a}. It should be noted that the
small-scale spectrum only needs to be steeper than~$k^{-3}$ for most
of the enstrophy to reside in the large scales. Hence, large-scale
structures may also be observed in simulations where a slope between
$-3$ and~$-5$ in the enstrophy range can be achieved, using a
large-scale dissipation (see next section). Also, if the spectrum on
the large scales is shallower than~$k^{-3}$, one may expect vortices
comparable in size to the forcing scale to form, because most of the
system's enstrophy is then distributed in that spectral region. This
has previously been noted by Paret and Tabeling \cite{Paret98}.  

In Section~\ref{constraints}, we show that (\ref{equilibrium}),
which may be rewritten as
\begin{eqnarray}
\label{equilibrium1}
\sum_{k<s}(s^2-k^2)k^2E(k) &=& \sum_{k>s}(k^2-s^2)k^2E(k), 
\end{eqnarray}
implies the spectral
exponents satisfy $\beta>5$ and $\alpha+\beta\ge8$, where $-\alpha$
($-\beta$) is the asymptotic slope of the range of wavenumbers lower
(higher) than the forcing wavenumber (in the usual case where the
forcing wavenumber is no greater than the geometric mean of the
integral and dissipation wavenumbers). For example, if the small-scale
spectrum is approximately $k^{-5}$, then the large-scale spectrum
should scale as $k^{-3}$. This would be consistent with the
observed $k^{-3}$ spectrum for the large-scale dynamics of the
atmosphere (Lilly \cite{Lilly83}). Now, a large-scale $k^{-3}$ spectrum
means that the enstrophy scales as $k^{-1}$; each octave in this
range contains approximately the same amount of enstrophy. Therefore,
the dissipation of energy is uniformly distributed among
successive octaves in the energy range, so that no inverse energy
cascade is possible. Nevertheless, a spectrum steeper than~$k^{-3}$ on
the large scales is allowed by (\ref{equilibrium1}). This is more
likely to occur if the small-scale spectrum is only marginally steeper
than $k^{-5}$ (see Section~\ref{constraints}). Unlike the KLB inverse
cascade, which carries virtually all of the injected energy to
ever-larger scales, this inverse cascade, if realizable, would only
carry a fraction of the energy input to the largest scales.

The dynamics of a bounded fluid in equilibrium is characteristically
different from the quasi-steady KLB picture. There is an infinite
energy reservoir at $k=0$ in unbounded systems that is forever
available to collect the energy transfer; this feature is absent in
bounded fluids (and in unbounded fluids satisfying the Poincar\'e
inequality).  Hence, the departure should not come as a surprise. What
seems ironic is that the enstrophy is only weakly dissipated in
bounded fluids at high Reynolds numbers but strongly dissipated in the
unbounded KLB case: in the bounded case the result $P/Z=s^2$
\cite{Tran02a} implies that the enstrophy dissipation rate $2\nu P/Z$
becomes $2\nu s^2$, while in the KLB theory it is approximately
$\nu k_\nu^2/\ln(k_\nu/s)$, which is much greater than $2\nu s^2$ since
$k_\nu\gg s$. 

There appears to be no simple generalization from the dynamics of a
bounded fluid in absolute equilibrium to that of its unbounded 
classical counterpart and {\it vice versa\/}. The familiar
reconciliation found in the literature is that the~$k^{-5/3}$ range is
modified or disrupted at the large scales when the inverse cascade
reaches the lowest available wavenumber.
If the KLB picture applies to non-equilibrium dynamics in a bounded
system, v.z. before the inverse cascade gets reflected by the spectral
boundary,\footnote{In a finite system, an inverse energy cascade
carries a non-negligible amount of enstrophy. If an inverse energy
cascade carrying virtually all of the energy input rate~$\epsilon$,
reaches a wavenumber $k_*$, it would transfer enstrophy at the
rate~$k_*^2\epsilon$ to wavenumber $k_*$. Hence, the ratio of  
the enstrophy accompanying the inverse energy cascade to the
enstrophy input is approximately given by $k_*^2/s^2$, independent of
the viscosity $\nu$.} then a dramatic adjustment of the spectrum has
to occur as an equilibrium is approached (see Smith and Yakhot
\cite{Smith94} for a discussion of the so-called finite-size
effects). We have not yet investigated in detail how this adjustment
takes place; however, it seems plausible that as the inverse cascade
hits the spectral boundary and gradually loses its strength
($dE/dt\rightarrow0$), a substantial amount of the energy gets bounced 
back to the forcing scale. As there is little dissipation
at the large scales, growth of the energy spectrum in the energy
range is inevitable. This growth proceeds until the large-scale energy
spectrum is sufficiently excited so that the enstrophy dissipation
occurs mainly in the vicinity of the forcing scale,
whereupon a forced-dissipative equilibrium is reached.
As $dE/dt\rightarrow0$, the enstrophy cascade (if initially present)
ceases since the quantity $P-s^2Z$ on the right-hand side of
(\ref{s2E-Z}) decreases to zero. As a consequence, a gradual
steepening of the enstrophy-range spectrum takes place (whatever the
spectral slope of the enstrophy range during the transient phase).
In Section~\ref{constraints} and Appendix~\ref{genslope}, we establish
that the sum of the steady-state spectral exponents in the energy and
enstrophy ranges must asymptotically approach $-8$. 

{\bf Remark 2.} It should be emphasized that the possibility of a
direct enstrophy cascade in a bounded fluid during the non-equilibrium
phase cannot be ruled out (not to say, however, that it actually occurs). 
For the case of a monoscale time-independent forcing, it is shown in
Tran and Shepherd \cite{Tran02a} that a direct enstrophy cascade is
not realizable on average, whether the average be taken on a chaotic
trajectory, limit cycle, or on the entire global attractor. This
result leaves only the possibility of a direct enstrophy cascade in a
neighborhood of the resulting monoscale stationary solution, on its
unstable manifold. In this region one simultaneously has $P-s^2Z>0$
and $s^2E-Z>0$ (see Tran and Shepherd \cite{Tran02a} and also Tran,
Shepherd, and Cho \cite{Tran02b}); the former inequality prevents
one from ruling out a direct enstrophy cascade. More quantitative
determination of the quantity $P-s^2Z$ (which is viscosity-dependent,
see \cite{Tran02b}) in this region will help assess the existence of the
enstrophy cascade (and how this would depend on the viscosity) as a
monoscale basic flow loses its stability.          

It is interesting to note that a $k^{-3}$ spectrum (or slightly
steeper), subject to experimental error, is observed in the laboratory
experiments of Paret, Jullien, and Tabeling \cite{Paret99}, and Rutgers
\cite{Rutgers98}. In these experiments, mechanical friction at the
bottom and top boundaries (in particular with the air, as noted in
\cite{Rutgers98}) of
the fluid could be sufficiently strong to outplay viscosity, so that
Ekman drag alone is essentially responsible for dissipation.
This might explain the observed spectra, according to the discussion in the
previous section. Moreover, the analysis of \cite{Tran02a} suggests that
a combination of strong Ekman drag and weak viscosity allows for such
a shallow spectrum to be realizable. Thus, there is no contradiction
between the observed spectra and the predicted $k^{-5}$ spectrum for a
Navier--Stokes fluid in equilibrium.   

\section{Constraints on constant spectral slopes}\label{constraints}

We now derive constraints on the spectral slopes for
bounded fluids in equilibrium and for a dual cascade in unbounded
fluids, particularly for a persistent inverse energy cascade.
We assume that the quasi-steady (steady) spectrum for the unbounded 
(bounded) case can be approximated by
\begin{eqnarray}
\label{spectrum}
E(k) &=& \cases{
ak^{-\alpha}&if $k_0 < k < s$,\cr
bk^{-\beta}&if $s < k < k_\nu$,\cr}
\end{eqnarray}
where $a,~b,~\alpha,~\beta$ are constants, $k_0$ is the lowest
wavenumber in the energy range ($k_0\rightarrow 0$ as
$t\rightarrow\infty$ for the unbounded case), and $k_\nu$ is the
highest wavenumber in the enstrophy range, beyond which the spectrum
is supposed to be steeper than $k^{-\beta}$. In
Appendix~\ref{genslope}, we show that the arguments below can even be
extended to the more realistic case, where the inertial-range slopes
depend on wavenumber. 

The quantity $P-s^2Z$ can then be estimated as
\begin{eqnarray}
\label{balanceA1}
\int_{k_0}^{\infty}(k^2-s^2)k^2E(k)\,dk 
&\ge&
a\int_{k_0}^s(k^2-s^2)k^{2-\alpha}\,dk +
b\int_s^{k_\nu}(k^2-s^2)k^{2-\beta}\,dk\nonumber\\ 
&=&
as^{5-\alpha}\int_{k_0/s}^1(\kappa^2-1)\kappa^{2-\alpha}\,d\kappa+
bs^{5-\beta}\int_{s/k_\nu}^1(1-\kappa^2)\kappa^{\beta-6}\,d\kappa\nonumber\\
&=&
as^{5-\alpha}\left(-\int_{k_0/s}^1(1-\kappa^2)\kappa^{2-\alpha}
\,d\kappa + \int_{s/k_\nu}^1(1-\kappa^2)
\kappa^{\beta-6}\,d\kappa \right),\nonumber\\
\end{eqnarray} 
where the inequality results from dropping the spectral contribution
beyond~$k_\nu$ (which is considerable if
$\beta\le5$), the second line is obtained by the respective changes of
variables $\kappa=k/s$ and $\kappa=s/k$ in the two integrals on the
right-hand-side of the first line, and the third line is obtained using the
continuity relation $as^{-\alpha}=bs^{-\beta}$.

In the bounded case the left-hand side of (\ref{balanceA1}) vanishes.  
Therefore, 
\begin{eqnarray}
\label{balanceA2}
\int_{k_0/s}^1(1-\kappa^2)\kappa^{2-\alpha}\,d\kappa
&\ge&
\int_{s/k_\nu}^1(1-\kappa^2)\kappa^{\beta-6}\,d\kappa.
\end{eqnarray} 
For a strong forcing at a relatively low wavenumber $s$, it is
reasonable to assume that $k_0/s\ge s/k_\nu$. This requires 
$2-\alpha\le\beta-6$, as the integrals in (\ref{balanceA2}) decrease
if the corresponding powers of $\kappa$ ($\beta-6$ and $2-\alpha$)
increase. That is, $\alpha+\beta\ge8.$

On the other hand, the convergence of the right-hand integral in
(\ref{balanceA2}), as $s/k_\nu\rightarrow 0$, requires also that
$\beta>5$. This result was derived in \cite{Tran02a}, on the basis that
the dissipation of enstrophy mainly occurs in the vicinity of the
forcing scale. Now if $\beta=5+\delta$, where it may be plausible that
$0<\delta\ll1$ for high Reynolds numbers, then $\alpha\ge3-\delta$. It
thus seems possible to obtain $\alpha\approx 3$. In the
limits $k_0/s\rightarrow0$ and $s/k_\nu\rightarrow 0$, as is usual for
high-Reynolds-number turbulence, the inequality $\alpha+\beta\ge8$
approaches an equality.  

{\bf Remark 3.} We caution that replacing the molecular viscosity 
$\nu k^2$ by a general viscosity $\nu_\mu k^{2\mu}$ ($\mu\ge 0$) in
the previous argument leads to the result $\alpha+\beta\ge
4+4\mu$. The significant dependence on $\mu$ of this constraint
suggests that the introduction of a hyperviscosity 
could seriously alter the expected steady-state spectral slopes. 

We now derive a condition for a persistent inverse energy cascade
in the unbounded case. We assume $\alpha < 3$, in accord with the
realization of an inverse cascade 
toward wavenumber $k=0$ that carries no enstrophy with it.
In the limits $k_0/s\rightarrow0$ and $s/k_\nu\rightarrow 0$,
the condition $P-s^2Z>0$ is guaranteed if $2-\alpha>\beta-6$ and
$\alpha < 3$, or equivalently, $\alpha+\beta<8$ and $\alpha < 3$. These
constraints admit a variety of spectra for a quasi-steady state in
which an inverse energy cascade to wavenumber zero, carrying with it
virtually no enstrophy, and a direct enstrophy cascade to the
dissipation wavenumber~$k_\nu$, carrying with it virtually no energy,
are allowed. Note that the inverse cascade scenario cannot be ruled
out even if $\beta>5$, i.e, in the absence of a direct enstrophy
cascade, provided that the inequality $\alpha+\beta<8$ holds. For the
KLB energy-range spectrum $k^{-5/3}$, it is interesting to note that
this condition requires only $\beta<19/3$. We therefore suggest that
an inverse energy cascade in the absence of a direct enstrophy cascade
can be realizable for a wide range of (modest) Reynolds numbers. 

{\bf Remark 4.} If the molecular viscosity $\nu k^2$ is replaced by a
hyperviscosity $\nu_\mu k^{2\mu}$ for $\mu>1$ in an unbounded fluid, 
the condition for a persistent inverse energy cascade $P-s^2Z>0$ is
replaced by $\int_0^\infty(k^2-s^2)k^{2\mu}E(k)\,dk>0$. This leads to
$\alpha+\beta<4+4\mu$. Of course, the condition $\alpha<3$ is required
for a zero-enstrophy-carrying inverse energy cascade. 

\section{Large-scale dissipation}

In this section, we examine how a large-scale
dissipation could be used to obtain a dual cascade, and in particular, a
direct enstrophy cascade. Consider~(\ref{E(k)evolution}) 
in the bounded case, with a general dissipation:
\begin{eqnarray}
\label{generalD(k)}
\frac{d}{dt}E(k) &=& T(k)-D(k)E(k)+F(k),
\end{eqnarray}
where $D(k)$ is a non-negative function of $k$.
Systems for which $D(k)$ vanishes in the intermediate wavenumber range,
including the forcing region, and for which the boundedness of energy is not
guaranteed (Eyink \cite{Eyink96}) have previously been studied in the
literature.

For a general $D(k)$, (\ref{s2E-Z}) becomes
\begin{eqnarray}
s^2\frac{d}{dt}E-\frac{d}{dt}Z &=& \sum_k(k^2-s^2)D(k)E(k),
\end{eqnarray}
which, in equilibrium, reduces to the balance equation
\begin{eqnarray}
\label{equilibrium3}
\sum_k(k^2-s^2)D(k)E(k) &=& 0.
\end{eqnarray}
A slightly different form of this equation is derived in \cite{Tran02a}.
Equation (\ref{equilibrium3}) implies that the energy-range spectrum 
is related to the enstrophy-range spectrum in an
intimate manner. For a given $D(k)$, an increase of the energy in one
range requires an increase of the energy in the other. Thus, a steeper
(shallower) energy-range spectrum corresponds to a shallower
(steeper) enstrophy-range spectrum, for fixed $E(s)$.
Another obvious consequence of~(\ref{equilibrium}) is that a nontrivial
equilibrium is not possible if $D(k)$ vanishes for all $k < s$.

We are interested in an expression for $D(k)$ that retains the usual
molecular viscosity and includes a large-scale dissipation. Thus, we
consider $D(k)=D_\ell(k)+2\nu k^2$, where $D_\ell(k)$ is a non-negative
function of $k$. This dissipation includes the physically relevant
case in which $D_\ell(k)$ is a positive constant representing friction 
from the planetary boundary layer in the geophysical context. Equation  
(\ref{equilibrium3}) then becomes
\begin{eqnarray}
\label{equilibrium4}
2\nu(P-s^2Z) &=& \sum_k(s^2-k^2)D_\ell(k)E(k). 
\end{eqnarray}
The left-hand side of (\ref{equilibrium4}) is the familiar term due 
to viscosity, which would vanish in the absence of the large-scale
dissipation. It can now become positive since the right-hand side
can be made positive in a variety of ways. Two popular forms of
$D_\ell(k)$ used in numerical simulations are the inverse viscosity
$D_\ell(k)=2\nu_\mu k^{2\mu}$, for $\mu<0$, and mechanical friction
restricted to the largest scales (say $k < k_\ell$): $D_\ell(k)\propto
H(k_\ell-k)$, where $H(k_\ell-k)$ is the Heaviside step function (see
Maltrud and Vallis \cite{Maltrud91}, for example). To facilitate the
formation of an enstrophy cascade, one might try to maximize the
right-hand side of (\ref{equilibrium4}). However, since we have no
{\it a priori\/} control over $E(k)$ for different $D_\ell(k)$, it is
not known how to maximize the product $D_\ell(k)E(k)$ for $k<s$. 
Nevertheless, restricting $D_\ell(k)$ to $k < s$ by setting
$D_\ell(k)=0$ for $k\geq s$ is reasonable since any non-zero
contribution to the right-hand side of (\ref{equilibrium4}) beyond $s$
is negative.      

If a positive value for the right-hand side of (\ref{equilibrium4})
can be achieved, the quantity $P-s^2Z$ will be large for
sufficiently small $\nu$. This may help break the constraint that the
enstrophy range slope must be steeper than $k^{-5}$ and allow for a
direct enstrophy cascade. An inverse cascade should be realizable, as
we expect most of the energy dissipation to occur at the large
scales. Thus, a dual cascade is possible. However, this cannot be
achieved without a cost, as~$D_\ell(k)$, which dissipates energy at
the rate $\sum_kD_\ell(k)E(k)$, also dissipates enstrophy on the large
scales at the rate $\sum_kD_\ell(k)k^2E(k)$. But
\begin{eqnarray}
k_0^2\sum_kD_\ell(k)E(k) &\leq& \sum_kD_\ell(k)k^2E(k),
\end{eqnarray}
where $k_0$ is the lowest wavenumber corresponding to the system
size. Hence, if $\sum_kD_\ell(k)E(k)$ is comparable to the
energy injection rate (which is ideally sought after in the spirit of
the KLB theory), then the ratio of the enstrophy dissipation rate at
the large scales to the enstrophy injection rate is greater than
$k_0^2/k_\max^2$. This fraction of the enstrophy dissipation at the
large scales may be small, but not negligible.
By allowing the enstrophy to be transferred to the large scales, a
non-negligible amount of enstrophy may be trapped in the forcing
region. If this is the case, then the spectrum in the forcing region
has to adjust dramatically (since $\nu$ is small and the large-scale
dissipation is assumed to be weak around the forcing scale) to balance
the trapped enstrophy. In Appendix~\ref{invviscslope}, we emphasize the
difficulty of obtaining an enstrophy-range spectrum shallower than
$k^{-5}$ with a large-scale dissipation that is well separated
from the forcing region.   

Numerical simulations of 2D turbulence can resolve up to a
certain wavenumber, say $k_T$. Therefore, there is always a finite
amount of energy dissipation at the small scales. This is analogous to
the dissipation of enstrophy at the large scales previously
considered, due to the finite size of the domain. If $D_h(k)$ represents
the small-scale dissipation coefficient, the enstrophy and energy
dissipations on the small scales are respectively given by
$\sum_kD_h(k)k^2E(k)$ and $\sum_kD_h(k)E(k)$. These quantities satisfy
\begin{eqnarray}
\sum_kD_h(k)k^2E(k) &\leq& k_T^2\sum_kD_h(k)E(k). 
\end{eqnarray}
Hence, if the dissipation of enstrophy by $D_h(k)$
is comparable to the enstrophy injection rate (which is ideally sought
after in the spirit of the KLB theory), then the ratio of the
energy dissipation rate at the small scales to the energy injection
rate is greater than $k_\min^2/k_T^2$. This is true for any $D_h(k)\ge0$,
including a hyperviscosity of arbitrary degree.

Intuitively, if a large-scale dissipation is extended to $s$, one would
expect it to absorb the reflected energy and keep the spectrum in the
forcing region from growing as $\nu\rightarrow0$. For a strong forcing and
strong large-scale dissipation $D_\ell(k)$ (confined to $k\le s$), it
may be hypothesized that the value of the right-hand side of
(\ref{equilibrium4}) is unaffected as $\nu\rightarrow0$, given all
else fixed. If this is the case, the quantity $P-s^2Z$ grows as
$\nu^{-1}$ and the ratio $P/Z$ is given by
\begin{eqnarray}
\frac{P}{Z} &=& s^2 + \frac{1}{2\nu Z}\sum_k(s^2-k^2)D_\ell(k)E(k). 
\end{eqnarray} 
This makes $P/Z\rightarrow\infty$ as $\nu\rightarrow0$, a favorable
limit for a direct enstrophy cascade, with a spectrum shallower than
$k^{-5}$. However, the resulting cascade would not have the physical
significance of the KLB theory since the direct enstrophy cascade
(regardless of the spectral slope) might only be marginal, with a
significant fraction of the enstrophy dissipated in the energy range
($k\le s$) due to the strong large-scale dissipation, contrary to the
classical theory.      

\section{CONCLUSION}

In this paper we have analysed the classical dual cascade theory
of 2D turbulence in unbounded fluids formulated by Kraichnan
\cite{Kraichnan67,Kraichnan71}, Leith \cite{Leith68}, and Batchelor
\cite{Batchelor69}. The main feature of the theory---the dual
cascade---is contrasted to the behaviour of 2D turbulence in a region
that satisfies the Poincar\'e inequality, such as a doubly periodic domain.
It is shown that the dual cascade picture, if realizable, would
strictly be an unbounded-system phenomenon. This important point is
not adequately stressed and has often led to confusion in the
literature. The familiar qualitative argument that the $k^{-5/3}$
range is modified or disrupted at the large scales when the inverse
energy cascade reaches the largest available scale in a bounded system
(assuming the applicability of the dual-cascade dynamics to the
transient phase) is inadequate. Two-dimensional turbulence either in a
doubly periodic domain or in an unbounded channel with a periodic
boundary condition on the across-channel dimension does not behave in
the manner predicted by KLB. In particular, the spectral slopes in
such systems are found to satisfy $\beta>5$ and $\alpha+\beta\ge8$,
where $-\alpha$ ($-\beta$) is the slope of the range of wavenumbers
lower (higher) than the forcing wavenumber. This result is well
supported by numerical simulations, which consistently find
enstrophy-range spectra steeper than the KLB prediction and
dynamically dominant large-scale structures (McWilliams
\cite{McWilliams84,McWilliams90}; Santangelo, Benzi, and Legras
\cite{Santangelo89}). It may even explain the observed large-scale
$k^{-3}$ spectrum in the atmosphere (Lilly \cite{Lilly83}, Boer and 
Shepherd \cite{Boer83}).

We have shown that a dual cascade in unbounded fluids is possible if
$\beta$ satisfies $3<\beta<5$; this includes both the classical
$\beta=3$ scaling as an extreme limit, and also the theories of
Saffman \cite{Saffman71}, Moffat \cite{Moffatt86}, and Sulem and
Frisch \cite{Sulem75}, which propose $\beta=4$, $\beta=11/3$, and
$\beta\le11/3$, respectively. Moreover, in the absence of a direct
enstrophy cascade ($\beta>5$), an inverse energy cascade,
corresponding to flows with low Reynolds numbers (which should be
relatively easy to simulate) cannot be ruled out. 

The fundamental difference between bounded and unbounded fluids is
that there is an infinite energy reservoir in the unbounded case,
which allows a persistent inverse energy cascade to ever-larger scales
to form, so that the energy eventually evades viscous dissipation
altogether. Provided that the spectrum near $k=0$ is shallower than
$k^{-3}$, the inverse cascade asymptotically carries no enstrophy.
This luxury is a consequence of both the unboundedness of the domain
(in both directions) and the scale-selectivity of the molecular 
viscosity. In addition to the simultaneous conservation of energy and
enstrophy, these properties constitute the basic building blocks of the KLB
theory. Another important hypothesis is the existence of a quasi-steady
state. As long as an inverse cascade is realizable and a quasi-steady
state can be established in an unbounded system, the cascade dynamics
are fundamentally distinct from what occurs in a bounded fluid in
equilibrium. There appears to be no sound basis for extending the
results from one case to the other. Of course, in an unbounded fluid
it is quite possible for an inverse energy cascade to exist without a
corresponding direct enstrophy cascade; in this case, there might then
be certain similarities between the dynamics of the bounded and
unbounded systems. 

The dissipation operator plays an important role in the spectral
distribution of energy. This is especially apparent in the balance
equation~(\ref{equilibrium3}) for a fluid in a doubly periodic domain:
the product of the energy spectrum and the spectral dissipation
function in the energy and enstrophy ranges are intimately
related. This information is important for numerical 2D turbulence
simulations, where various dissipation mechanisms are employed: it
should help researchers rule out certain spectral distributions and
anticipate possible outcomes for a given dissipation mechanism. 
Finally, we showed that a large-scale dissipation could give rise to a
direct enstrophy cascade since the quantity $P-s^2Z$ could grow as
$1/\nu$, until $P/Z\gg s^2$.  

\appendix
\section{Constraints on general spectral slopes}\label{genslope}

Strictly speaking, the spectrum~(\ref{spectrum}) is too
simplistic; actual spectral slopes will tend to vary monotonically with
wavenumber (particularly in the enstrophy range, as one approaches the
onset of the dissipation range). The arguments of Section~\ref{constraints}
can be readily extended to more general spectra. Normalizing all
wavenumbers so that $s=1$, we express the energy spectrum as 
\begin{eqnarray}
E(k)=a \cases{ k^{-\alpha(k)}&if $k_0 < k < 1$,\cr
k^{-\beta(k)}&if $1 \le k < k_1$,\cr
k^{-\gamma(k)}&if $k_1\le k \le k_T$,\cr}
\end{eqnarray}
with $\beta(k_1)=\gamma(k_1)$, where $k_0$ is the lower spectral cutoff
wavenumber (determined by the domain size), $k_1$ is the highest wavenumber
in the enstrophy range, and $k_T$ is the highest retained (truncation)
wavenumber.
\def\supa{\overline\alpha}
\def\supb{\overline\beta}
Equation~(\ref{equilibrium}), or equivalently~(\ref{equilibrium3}),
then appears as
\begin{eqnarray}
\int_{k_0}^1(1-k^2)k^{2-\alpha(k)}dk &=&
\int_1^{k_1}(k^2-1)k^{2-\beta(k)}dk+
\int_{k_1}^{k_T}(k^2-1)k^{2-\gamma(k)}\,dk.
\end{eqnarray}
The change of variable $\kappa=1/k$ in the integrals on the right-hand
side yields
\begin{eqnarray}
\label{case1}
\int_{k_0}^1(1-\kappa^2)\kappa^{2-\alpha(\kappa)}
d\kappa &=&
\int_{1/k_1}^1(1-\kappa^2)\kappa^{\beta(1/\kappa)-6}d\kappa
+\epsilon_\nu,
\end{eqnarray}
where $\epsilon_\nu=\int_{1/k_T}^{1/k_1}(1-\kappa^2)
\kappa^{\gamma(1/\kappa)-6}\,dk \ge 0$.
For bounded turbulence, we restrict our attention to the usual case
where $k_1 \ge 1/k_0$. Since
$\int_{1/k_1}^1(1-\kappa^2)\kappa^{\theta}d\kappa$ is a strictly
decreasing function of $\theta$, we find that the maximum slopes 
$\supa=\displaystyle\sup_k \alpha(k)$ and $\supb=\displaystyle\sup_k
\beta(k)$ satisfy 
\begin{eqnarray}
\int_{1/k_1}^1(1-\kappa^2)\kappa^{2-\supa}\,d\kappa
&\ge&\int_{1/k_1}^1(1-\kappa^2)\kappa^{2-\alpha(\kappa)}\,d\kappa
\ge\int_{k_0}^1(1-\kappa^2)\kappa^{2-\alpha(\kappa)}\,d\kappa\nonumber\\
&\ge&\int_{1/k_1}^1(1-\kappa^2)\kappa^{\beta(1/\kappa)-6}\,d\kappa
\ge\int_{1/k_1}^1(1-\kappa^2)\kappa^{\supb-6}\,d\kappa.\nonumber
\end{eqnarray}
Hence $2-\supa\le\supb-6$; that is, $\supa+\supb\ge 8.$
\def\infa{\underline\alpha}
\def\infb{\underline\beta}

One can also obtain estimates for the minimum slopes
$\infa=\displaystyle\inf_k \alpha(k)$ and
$\infb=\displaystyle\inf_k \beta(k)$,
assuming $\gamma(k)\ge\infb $:
\begin{eqnarray}
\int_{k_0}^1(1-\kappa^2)\kappa^{2-\infa}\,d\kappa
&&\le\int_{k_0}^1(1-\kappa^2)\kappa^{2-\alpha(\kappa)}\,d\kappa
=\int_{1/k_1}^1(1-\kappa^2)\kappa^{\beta(1/\kappa)-6}\,d\kappa+
\epsilon_\nu\nonumber\\
&&\le\int_{1/k_T}^1(1-\kappa^2)\kappa^{\infb-6}\,d\kappa.
\label{infbalance}
\end{eqnarray}
In the asymptotic limit as $k_0\rightarrow 0$ and 
$k_T\rightarrow\infty$, we deduce $\infa+\infb \le 8.$

It is instructive to specialize these results to the
case of constant slopes, where $\infa=\supa=\alpha$ and
$\infb=\supb=\beta$  (e.g. if $k_1 \ll k_\nu$).
For a bounded fluid satisfying $k_1\ge 1/k_0$, we then see that
$\lim_{k_0\rightarrow 0} \alpha+\beta=8$. 

\section{Spectral slopes for a system with an inverse viscosity}
\label{invviscslope}

It was suggested in Section~4 that a large-scale dissipation, well
separated from the forcing scale, may not give rise to the desired
direct enstrophy cascade and the corresponding $k^{-3}$ spectrum. To
demonstrate this point, we consider the special form
$D(k)=2\nu'k^{-2}+2\nu k^2$, with $\nu's^{-2}=\nu s^2$,
so that at the forcing scale, the inverse viscosity and the usual
molecular viscosity have the same strength. 
Equation (\ref{equilibrium3}) becomes 
\begin{eqnarray}
\sum_k(k^2-s^2)(\nu'k^{-2}+\nu k^2)E(k) &=& 0. 
\end{eqnarray}
Assume the spectral scaling (\ref{spectrum}) and follow the
steps leading to (\ref{balanceA2}), we obtain 
\begin{eqnarray} 
\label{balanceB1}
\int_{k_0/s}^1(1-\kappa^2)(\kappa^{-2}+\kappa^2)\kappa^{-\alpha}\,d\kappa 
&\ge& 
\int_{s/k_\nu}^1(1-\kappa^2)(\kappa^2+\kappa^{-2})\kappa^{\beta-4}\,d\kappa.
\end{eqnarray}  
In the case $k_0/s\ge s/k_\nu$ it follows that $\alpha+\beta \ge 4$.
In the limit $s/k_\nu\rightarrow 0$, we still require $\beta>5$.
For the KLB enstrophy-range spectrum to be realizable it is necessary that
$\alpha\ge 1$; moreover, $s/k_\nu$ cannot be much smaller than $k_0/s$. The
former condition does not seem to be plausible in the presence of an inverse
viscosity, while the latter condition requires an unphysically narrow
enstrophy range for a forcing at relatively small wavenumbers.   

{\bf Acknowledgements}

We would like to thank two anonymous referees for their
constructive comments, which helped us clarify and improve
the manuscript. This work was supported by the Natural Sciences and
Engineering Research Council of Canada. CVT was also supported by a
Pacific Institute for the Mathematical Sciences Postdoctoral Fellowship.
 


\end{document}